\documentclass[12pt]{article}
\usepackage{epsfig}
\begin{document}

\begin{titlepage}

\begin{center}

\hfill{FTUV-04-1209}

\hfill{IFIC-04-70}

\vspace{2cm}

{\Large  \bf Triquark correlations and pentaquarks in a QCD sum
rule approach} \\

\vspace{0.50cm}

\renewcommand{\thefootnote}{\fnsymbol{footnote}}
Hee-Jung Lee $^{a}$\footnote{Heejung.Lee@uv.es},
N.I.Kochelev$^{b}$\footnote{kochelev@thsun1.jinr.ru}, V. Vento$^a$
\footnote{Vicente.Vento@uv.es}

\vspace{0.50cm}
{(a) \it Departament de F\'{\i}sica Te\`orica and Institut de F\'{\i}sica Corpuscular,\\
Universitat de Val\`encia-CSIC, E-46100 Burjassot (Valencia), Spain} \\

\vskip 1ex

{(b) \it Bogoliubov Laboratory of Theoretical Physics,\\
Joint Institute for Nuclear Research, Dubna, Moscow region, 141980
Russia}

\end{center}

\vskip 0.5cm

\centerline{\bf Abstract}

The role of quark correlations in the description of hadron
dynamics in many domains of physics, from low energy dynamics to
very hot(dense) systems, is being appreciated. Strong correlations
of two quarks (diquark) have been widely investigated in this
respect. Recently, we have proposed a dynamical scheme to describe
the $\Theta^+$ pentaquark in which also three quark correlations
(triquark) were instrumental in producing a low mass exotic state.
We perform a study, within the QCD sum rule approach including OPE
and direct instanton contributions, of triquark correlations and
obtain two quasi-bound light $ud\bar{s}$ color quark clusters of
800 MeV and 930 MeV respectively.

\vskip 0.3cm \leftline{Pacs: 12.38.Aw, 12.38.Lg, 12.39.Ba, 12.39.-x}

\leftline{Keywords: quarks, instanton, hadrons, pentaquark}

\vspace{1cm}

\end{titlepage}

\section{Introduction}

In 2003 evidence was reported of a very narrow exotic baryon of
mass $\approx$ 1540 MeV and small width \cite{evidence0}. This so
called $\Theta^+$ pentaquark with a minimal quark content
$uudd\bar{s}$ has motivated tremendous experimental and
theoretical activity since its first sightings. Many experiments
confirmed the observation \cite{evidence1} and new exotics where
reported \cite{evidence2}. Theorist have aimed at understanding
these states from the point of view of known low energy
realizations of QCD \cite{chiral,bag,jw,kl,klv,theory}. With time
the situation has become confusing \cite{negativetheory}. The
experimental status of $\Theta^+$ is controversial since several
experiments have reported searches with negative results and
moreover no single experiment has confirmed the heavier
exotics\cite{negativeexperiment}. What seems to be established is
that if these states exist they probe special features of QCD
dynamics which will explain their rarity. Even if they do not
exist, the efforts thus far have discovered dynamical features of
QCD which favor clustering and which might be useful at higher
densities and/or temperature.

The pentaquark arises quite naturally in chiral soliton schemes
\footnote{Some authors have questioned the consistency of this
calculation~\cite{cohenpoby}.}\cite{chiral}. However if one uses
quark degrees of freedom the multiparticle nature of the state
makes the dynamical analysis more elaborate. Conventional dynamics
\cite{bag} leads to exotic baryons which are to heavy and their
widths too large. Cluster schemes have been proposed which tend to
explain the data. Jaffe and Wilczek  rely on strong diquark
correlations and Pauli blocking to generate a low mass, small
width, state \cite{jw}. Lipkin and Karliner propose a
triquark-diquark system induced by a generalized color magnetic
interaction \cite{kl}. In our scheme \cite{klv}, with the
clustering of Karliner and Lipkin, the one gluon exchange (OGE)
interaction plays a minor role and the correlations are built in
by a strong Instanton Induced Interaction (I3)
\cite{shuryak,diakonov}. The specific feature of the I3 leading to
clustering  is its strong flavor and spin dependence, i.e., due to
the Pauli Principle of the quarks in the zero modes of the
instanton field the interaction is only non vanishing  between
different quark flavors. The strength of the instanton induced
attraction in the scalar-isoscalar diquark channel is enough to
produce an almost bound color state \cite{shuryak}.

In ref.\cite{klv} we presented arguments in favor of the formation
inside the pentaquark, due to the coupling of the instanton field,
of a light color cluster with flavor content $ud\bar s$ . In such
a system a strong attraction is possible not only in the
quark-quark, but also in quark-antiquark channel. Therefore, the
feasibility for formation of  a light, $\approx 750$ MeV, triquark
state was discussed. In order to confirm, from a more fundamental
point of view, the results of our model calculation we proceed to
use the QCD sum rule (SR) approach. As emphasized, in our model
calculation, due to the particular spin-flavor-color structure of
the pentaquark wave function, new types of two- and three-body I3s
between the quarks, different from those appearing in conventional
hadrons, are possible. Therefore, the analysis of the instanton
effects on the properties of the multiquark hadrons within the SR
approach is an interesting and actual problem, which describes new
types of quark-quark correlations.

We present the first calculation of instanton effects in the
multiquark sector of QCD within the QCD sum rule approach. Our
considerations will start from the discussion of the direct
instanton contribution to the sum rules for the nucleon and
thereafter we will calculate instanton effects on the mass of a
colored $ud\bar s$ triquark.

\section{QCD sum rule approach for the nucleon and triquark states}

The study of correlations using SR is not new. A $ud$-diquark
color system was considered within the QCD sum rule approach
including instanton contributions \cite{shuryak3,faccioli} and it
was shown that instanton induced attraction leads to a bound state
for the isoscalar diquark with mass $m_{ud}\approx 420\sim 450$
MeV. Our model study  \cite{klv} of the $\Theta^+$ has discovered
the possibility of physically relevant triquark correlations.
Never mind the existence of the pentaquark, put in jeopardy by the
last experimental analysis, the study of all kinds of quark
correlations is an interesting project in itself, because they
maybe important dynamical mechanisms in various domains of
physics.

One important conceptual distinction between the study of physical
hadrons and color correlations using sum rules is the fact that
the latter are not color singlets, and therefore are not physical
states. The way to proceed is to build a color singlet current
adding a sterile quark (antiquark) to them. In particular, a
problem that has been discussed in detail is gauge independence in
the extraction of their masses from the SR \cite{shuryak3}. There,
it is argued that one can consider colorless currents for the
diquark (triquark) with an additional heavy quark (antiquark) and
in this way avoid the problem of gauge invariance. However, since
the heavy quarks (antiquarks) interact very weakly with the
instantons, our results below will not change significantly.

The main object in the SR approach, based on operator product
expansion (OPE), is the correlator of two interpolating currents,
with the quantum numbers of the particle under scrutiny, and which
is given by

\begin{equation}
\Pi(p^2)=i\int d^4x \ e^{ip\cdot x}\langle 0| T
\eta(x)\bar{\eta}(0)|0\rangle. \label{corr}
\end{equation}

For spin 1/2 baryons, the correlator can be decomposed into two
functions
\begin{equation}
\Pi(p^2)=\hat{p}\Pi_1(p^2)+\Pi_2(p^2).
\end{equation}
The spectral representation for the imaginary part of the
correlator entering the dispersion relation is of the form
\begin{equation}
Im \Pi(s)=\pi |\lambda_B|^2 (\hat{p} + M_B) \delta(s^2-M_B^2)
+\pi\theta(s^2-s_0^2)(\hat{p} Im \Pi_1(s^2)+ Im \Pi_2(s^2)),
\end{equation}
where $M_B$ and $\lambda_B$ are the mass and coupling strength of
the ground state onto which the current projects, and $s_0$ is a
threshold, which will be used below to relate the properties of
the nucleon and color triquark state with the OPE and the direct
instanton contributions to the correlator.

We will consider the OPE contributions to the correlator up to the
dimension six. The diagrams shown in Fig.~1 represent the OPE
contributions to $\Pi_1$ for the nucleon. The diagrams in Fig.~2
contain contributions for non vanishing quark mass to $\Pi_2$  and
therefore can be applied to the nucleon and to the triquark state.
We will only use $\Pi_2$ to calculate the properties of the
triquark state.

\begin{figure}[htb]
\centerline{\epsfig{file=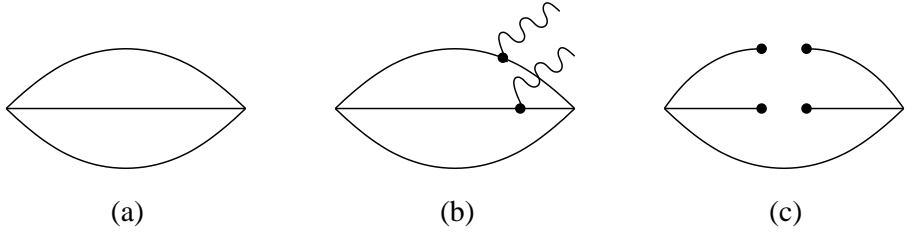,width=12cm,angle=0}}\
\caption{Diagrams entering the calculation of  $\Pi_1(p^2)$ for
the nucleon in the QCD sum rule OPE approach.}
\end{figure}

\begin{figure}[htb]
\centerline{\epsfig{file=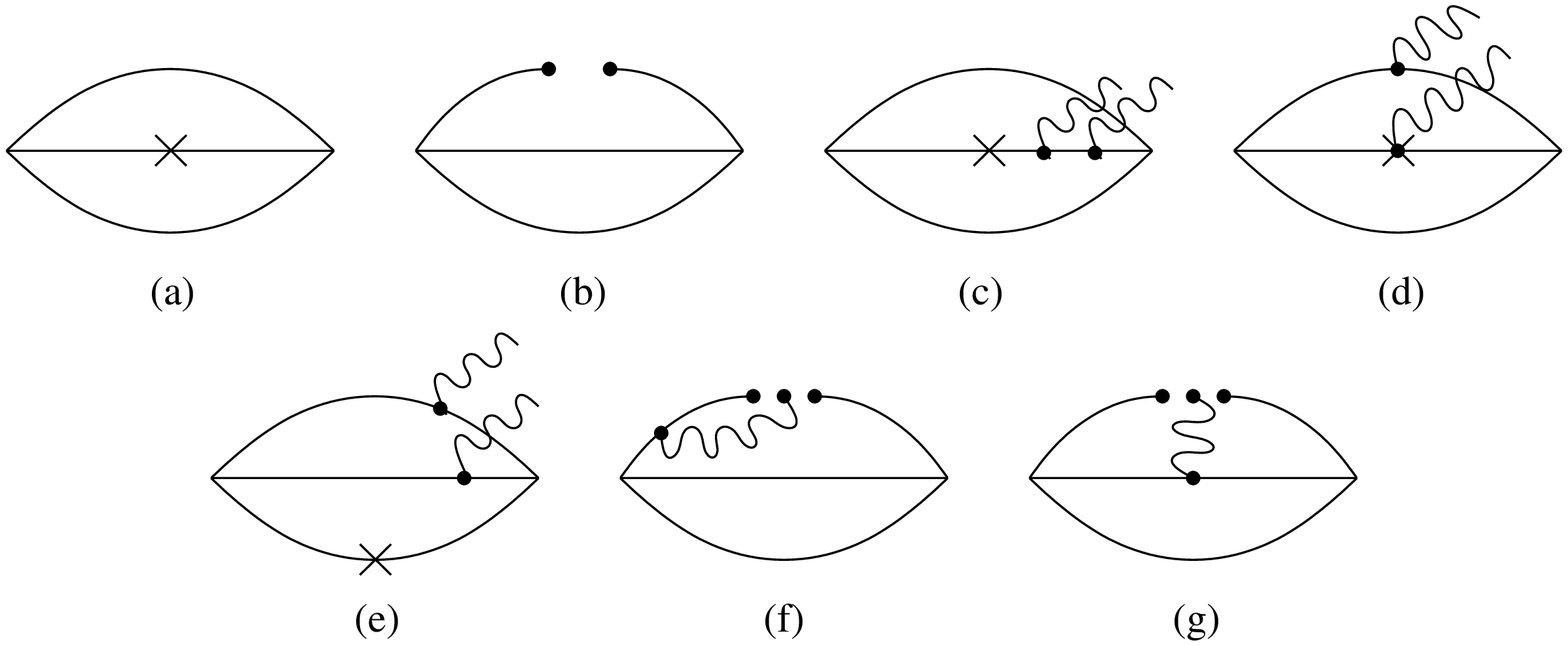,width=12cm,angle=0}}\
\caption{Diagrams entering the calculation of $\Pi_2(p^2)$ in the
SR OPE approach. The insertion $\times$ denotes the quark mass.}
\end{figure}

The quark propagator to this order, has the form
\begin{eqnarray}
S^q_{ab}(x)&=&-i\langle 0|Tq_a(x)\bar{q}_b(0)|0\rangle\nonumber\\
&=&\delta_{ab}(\hat{x}F_1^q+F_2^q)-i\tilde{g}G^{\mu\nu}_{ab}\frac{1}{x^2}
(\hat{x}\sigma_{\mu\nu}+\sigma_{\mu\nu}\hat{x})
-m_q\tilde{g}G^{\mu\nu}_{ab}\sigma_{\mu\nu}\ln(-x^2),
\end{eqnarray}
where $a,b$ are the color indices and ${\tilde g}=g_c/32\pi^2$.
The two functions entering the propagator are given by
\begin{eqnarray}
F_1^q&=&\frac{1}{2\pi^2 x^4}+\frac{m_q \langle
\bar{q}q\rangle}{48}
+i\frac{m_qx^2}{2^7\cdot3^2}g_c\langle\bar{q}\sigma\cdot G
q\rangle
\nonumber\\
F_2^q&=&i\frac{m_q}{4\pi^2x^2}+i\frac{\langle \bar{q}q\rangle}{12}
-\frac{x^2}{192}g_c\langle\bar{q}\sigma\cdot G q\rangle
+i\frac{m_qg_c^2}{2^9\cdot 3\pi^2}\langle G^2\rangle x^2\ln(-x^2)
\end{eqnarray}
where $\langle{\cal O} \rangle$ denotes the vacuum condensate of
the operator.

The gluon condensate and the mixed condensate are defined by
\begin{equation}
\langle G^2 \rangle=\langle G_{\mu\nu}G^{\mu\nu}\rangle,\ \
\langle\bar{q}\sigma\cdot G
q\rangle=\langle\bar{q}\sigma_{\mu\nu}\cdot G^{\mu\nu} q\rangle
\end{equation}
where $\sigma_{\mu\nu}$ is defined by
\begin{equation}
\sigma_{\mu\nu}=\frac{1}{2}[\gamma_\mu,\gamma_\nu] \ .
\end{equation}
We will assume that the current masses for the  $u, d$ quarks are
zero, and therefore $F^u_1=F^d_1,\ F^u_2=F^d_2.$

Let us begin by discussing the nucleon. The positive parity
interpolating current is
\begin{equation}
\eta_{tot}(x)=f\eta(x)+t\eta_1(x),
\end{equation}
where $f$ and $t$ are two real parameters characterizing the
mixing between the currents which are
\begin{eqnarray}
\eta^N(x) &=&\epsilon^{abc}[u^T_a(x)\Gamma d_b(x)]u_c(x)
\nonumber\\
\eta^N_1(x) &=&\epsilon^{abc}[u^T_a(x)\Gamma_1
d_b(x)]\gamma_5u_c(x),
\end{eqnarray}
with $\Gamma=C\gamma_5, \Gamma_1=C $.

Using the conventional SR formalism after performing the Borel
transform, we obtain two OPE sum rules for nucleon,
\begin{eqnarray}
&&\frac{1}{4}(5t^2+2tf+5f^2)E_2M^6+\frac{1}{16}(5t^2+2tf+5f^2)bE_0M^2
\nonumber\\
&&\hspace{4.7cm}
+\frac{2}{3}(7f^2-2tf-5t^2)a^2=\; \tilde{\lambda}_N^2\; e^{-M_N^2/M^2}\ ,\\
&&(7f^2-2tf-5t^2)aE_1M^4-3(f^2-t^2)m_0^2aE_0M^2\;\;\; =
\tilde{\lambda}_N^2M_N e^{-M_N^2/M^2}.
\end{eqnarray}

Let us apply the same formalism to the  $ud\bar s$
triquark state with isospin $I=0$, spin $S=1/2$ and color
$C=3_c$  in the pentaquark. We will consider two types of
triquark states with different color structure for the
$ud$ subsystem labelled A, $C_{ud}=\bar 3_c$  and B,
$C_{ud}= 6_c$ (see \cite{klv}).

The A state  has a non vanishing overlap with the currents
\begin{eqnarray}
\eta^A&=&\frac{1}{4}\epsilon_{abc}\epsilon_{bde}[u^T_d\Gamma
d_e]C\bar{s}_c^T, \ \
\eta^A_1=\frac{1}{4}\epsilon_{abc}\epsilon_{bde}[u^T_d\Gamma_1
d_e] \gamma_5C\bar{s}_c^T, \label{tricurrentA}
\end{eqnarray}
which correspond to the  mixture of scalar and pseudoscalar
isosinglet ($ud$) diquark appearing in its wave function.

The B state has a non vanishing overlap with the current
\footnote{There is another current defined in terms of the tensor
$\sigma_{\mu \nu}$ which is not relevant for the purposes of the
present calculation.}
\begin{equation}
\eta^B(x)=\frac{1}{4\sqrt{3}}[u_a^T(x)C\gamma_\mu
d_b(x)+u_b^T(x)C\gamma_\mu d_a(x)] \gamma_5\gamma^\mu
C\bar{s}_b^T.\label{tricurrentB}
\end{equation}
All these interpolating currents are of negative parity.

For the triquark states only chirality odd SR will be considered,
because, as will be shown later, only chirality odd SR have a good
stability plateau when direct instanton contributions are
incorporated.

Thus, for the A state we have
\begin{eqnarray}
&&(f^2-t^2)\bigg(\frac{m_s}{6}E_2M^6+\frac{f_s}{6}aE_1M^4
-\frac{f_s}{12}am_0^2E_0M^2
\nonumber\\
&&\hspace{2cm}-\frac{m_s}{24}bE_0M^2(v(M^2)-1/2) \bigg)
=\tilde{\lambda}_A^2M_Ae^{-M_A^2/M^2},
\end{eqnarray}
and for the B state the result is
\begin{eqnarray}
& &\frac{2}{9}m_sE_2M^6+\frac{2}{9}f_s aE_1M^4
+\frac{1}{36}f_sm_0^2 aE_0M^2
\nonumber\\
&&\hspace{1.3cm}+\frac{1}{72}m_sbE_0M^2(v(M)+1)
=\tilde{\lambda}_B^2M_Be^{-M_B^2/M^2},
\end{eqnarray}
where  $v(M)=\ln(M^2\rho_c^2/4)+\gamma_{EM}-1$.

Above we considered contributions to the correlator up to
dimension six and up to orders ${\cal O}(m_s)$, ${\cal O}(m_s
g_c)$, ${\cal O}(m_s g_c^2)$, and did not consider ${\cal
O}(\alpha_s)$ corrections. The parameters are introduced by the
following relations
\begin{eqnarray}
&&\langle \bar{u} u\rangle=-\frac{a}{(2\pi)^2}\ ,\ \ g_c^2\langle
G\cdot G\rangle=b,\ \ ig_c\langle\bar{u}\sigma\cdot
Gu\rangle=m_0^2\langle \bar{u} u\rangle\ ,\ \
\nonumber\\
&&\hspace{1.5cm} \frac{\langle \bar{s} s\rangle}{\langle \bar{u}
u\rangle}= \frac{\langle\bar{s}\sigma\cdot G
s\rangle}{\langle\bar{u}\sigma\cdot G u\rangle}=f_s\ , \ \ \
\tilde{\lambda}_B=(4\pi)^2\lambda_B\ ,
\end{eqnarray}
and the  needed functions $E_n$ are given by
\begin{equation}
E_n(x)=1-e^{-x}\sum_n\frac{x^n}{n!}\ ,\ \ \ {\rm with}\ \ \
x=\frac{s_0^2}{M_B^2}\ .
\end{equation}

We will use the following values for parameters \cite{ioffe}
\begin{eqnarray}
&&\langle\bar q q\rangle=-(250\ {\rm MeV})^3, \  b=0.24\ {\rm
GeV}^4, \
m_0^2=1\ {\rm GeV}^2, \nonumber\\
&& m_s=150\ {\rm MeV}, \  f_s=0.8\ . \nonumber
\end{eqnarray}

\section{QCD sum rules  with direct instanton contributions}

In the OPE based SR for the nucleon, the contributions due to
large size vacuum fluctuations of quark and gluon fields have been
taken into account (Fig.1 and 2).  If only such fluctuations are
important in determining the mass of a particle, with given
quantum numbers, then the OPE is valid and one can safely restrict
the calculation to a finite number of terms in the expansion.
However, in the QCD vacuum, there are strong fluctuations of small
size associated with the gluon fields, namely the instantons,
which can lead to a significant modification of the OPE QCD sum
rules \cite{shuryak2,DEKS,dorkoch,forkel}. For example, the
instantons can produce a large violation of factorization in some
four- quark vacuum-vacuum matrix elements and can lead to the
appearance of additional exponential terms in the current
correlator which have nothing to do with the standard power-like
$1/q^{2n}$ of the OPE expansion.

We proceed to incorporate the instanton contributions in our
calculation. To do so we have to have to rotate all our equations
to Euclidian space-time, where the instantons are defined,
according to $\hat{x}_M=-i\hat{x}_E,\ x^2_M=-x^2_E,$ and
$\langle\bar{q}q\rangle_M=-i\langle\bar{q}q\rangle_E$.

The propagator has two terms, the standard one ($st$) and the one
associated to the instanton contributions ($inst$),
\begin{equation}
S^q_{ab}(x,y)=S^{q,st}_{ab}(x,y)+S^{q,inst}_{ab}(x,y),
\label{prop}
\end{equation}

For the standard quark propagator $S^{q,st} $ we use the free
propagator with mass and quark condensate corrections, i.e.,
\begin{eqnarray}
S^{q,st}_{ab}(x,y)&=&\delta_{ab}\bigg(\frac{\hat{x}-\hat{y}}{2\pi^2(x-y)^4}
+i\frac{m_q}{4\pi^2(x-y)^2}+i\frac{\langle\bar{q}q\rangle}{12}\bigg)
\nonumber\\
&&\longrightarrow
\delta_{ab}\bigg(-i\frac{(\hat{x}-\hat{y})_E}{2\pi^2(x-y)^4}
-i\frac{m_q}{4\pi^2(x-y)^2}+\frac{\langle\bar{q}q\rangle_E}{12}\bigg)
\end{eqnarray}
The leading effect of instantons is provided by the zero quark
mode approximation which leads to the following ansatz for the
quark propagator in the instanton background, \cite{shuryak}
\begin{equation}
S^{q,inst}_{ab}(x,y)=A_q(x,y)[(\hat{x}-\hat{z}_0)\gamma_\mu\gamma_\nu
(\hat{y}-\hat{z}_0)(1-\gamma_5)] (U\tau^-_\mu\tau^+_\nu
U^\dagger)_{ab}
\end{equation}
where $$ A_q(x,y)=-i\frac{\rho^2}{16\pi^2
m_q^{*}}\phi(x-z_0)\phi(y-z_0) $$ and
$$\phi(x-z_0)=\frac{1}{\sqrt{(x-z_0)^2}[(x-z_0)^2+\rho^2]^{3/2}}\
.$$ Note that $\rho$ stands for the instanton size, $\rho_c$ is
the average instanton size and $z_0 $ the center of the instanton;
$U$ represents the color orientation matrix of the instanton in
$SU(3)_c$ and $\tau_{\mu,\nu}$ are $SU(2)_c$ matrices;
$m_q^*=m_{cur}^q-2\pi^2\rho_c^2\langle\bar q q\rangle/3$ is the
effective quark mass in the instanton vacuum and $m_{cur}^q$ the
current quark mass. The final result should be multiplied by a
factor of two, due to the antiinstanton contribution, and has to
be integrated over the instanton density $\int n(\rho)d\rho$
\footnote{For a discussion on the possible values of the
parameters of  this instanton model see ref. \cite{shuryak3}.}.

An important selection rule for the quarks in the instanton field
reads
\begin{equation}
\stackrel{\rightarrow}{\sigma_i}\bigoplus \vec \tau_i=0,
\label{rule}
\end{equation}
where $\sigma_i$ is usual spin and $\tau_i$ is color spin of the
quark. This selection rule leads to the vanishing of the instanton
two-body quark contribution to masses of particles from the baryon
decuplet and forbids also the three-body instanton induced
interaction to the colorless baryons. In Fig.3 the two-body and
three-body instanton induced contributions to current correlator
are shown.


\begin{figure}[htb]
\centering \epsfig{file=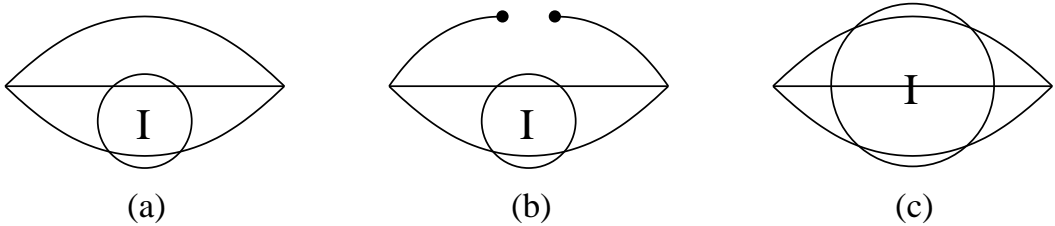,width=14cm} 
\caption{ The  a) two-body instanton induced contribution to
$\Pi_1$, b) two-body instanton contribution to $\Pi_2$ and c)
three-body instanton contribution to $\Pi_2$. In the figure $I$
denotes the instanton.}
\end{figure}


Using a model for instanton density defined by
$n(\rho)=n_{eff}\delta(\rho-\rho_c)$ \cite{shuryak2}, we calculate
the instanton contributions to the correlator of the nucleon
current, which after Borel transformation are given by
\begin{eqnarray}
\Pi_1^N(M) &=&\frac{3n_{eff}(f^2-t^2)}{4^3\pi^2\rho_c^4(m_u^*)^2}
\bigg[\frac{64}{5}\bigg(1-\frac{24}{7z^2}\bigg)
\nonumber\\
&&+\frac{z^8}{4}\int_0^1dy\frac{1}{y^2(1-y)^2}
\frac{(X^3+6X^2+18X+24)e^{-X}}{X^5}\bigg]\label{nucleon1}
\\
\Pi_2^N(M) &=&-\frac{n_{eff}\langle
\bar{q}q\rangle_M\rho_c^4}{3\cdot 2^6(m_u^*)^2}
(13t^2+10tf+13f^2)M^6e^{-z^2/2}\nonumber \\
&&\times \bigg(K_0(z^2/2)+K_1(z^2/2)\bigg), \label{nucleon2}
\end{eqnarray}
where $z=M\rho_c$ and $X=z^2/(4y(1-y))$.

The instanton contribution to the color $ud\bar s$ states has
a more complicated structure. The two-body instanton effects
to $\Pi_2$ for the correlator of state A is given, in
configuration space, by
\begin{eqnarray}
\langle T\eta^A_{tot}(x)\bar{\eta}^A_{tot}(0)\rangle^{2}
&=&\frac{in_{eff}\rho_c^4}{2\pi^4}
\frac{1}{z_0^2(x-z_0)^2[(x-z_0)^2+\rho_c^2]^3[z_0^2+\rho_c^2]^3}
\\
&& \bigg\{\frac{z_0^2(x-z_0)^2}{3m_u^*m_d^*}
\bigg[(t^2+f^2)\bigg(\frac{\langle\bar{s}s\rangle_E}{12}
-i\frac{m_s}{4\pi^2x^2}\bigg)-(t^2-f^2)\frac{i\hat{x}}{2\pi^2x^4}\bigg]
\nonumber\\
&&+(t+f)^2\frac{\langle\bar{q}q\rangle_E}{4^3\cdot6m_u^*m_s^*}
\bigg(4((x-z_0)\cdot z_0)^2-\frac{4}{3}((x\cdot
z_0)^2-x^2z_0^2)\bigg)\bigg\}\ , \nonumber
\end{eqnarray}
where we have used the assumption
$\langle\bar{u}u\rangle=\langle\bar{d}d\rangle$ and denoted them
by $\langle\bar{q}q\rangle$, while the three-body contribution is
given by
\begin{eqnarray}
\langle T\eta^A_{tot}(x)\bar{\eta}^A_{tot}(0)\rangle^{3}
&=&-(t+f)^2\frac{n_{eff}\rho_c^6}{12\pi^6m_u^*m_d^*m_s^*}
\nonumber\\
&&\times\frac{(x-z_0)\cdot z_0}
{\sqrt{(x-z_0)^2}\sqrt{z_0^2}[(x-z_0)^2+\rho_c^2]^{9/2}[z_0^2+\rho_c^2]^{9/2}}\
\label{tricurrent}
\end{eqnarray}

The two-body instanton effects to $\Pi_2$ for the correlator of
state B is given, in configuration space, by
\begin{equation}
\langle T\eta^B(x)\bar{\eta}^B(0)\rangle^{2}
=i\frac{11n_{eff}\rho_c^4\langle\bar{q}q\rangle_E}{4\cdot108\pi^4m_q^*m_s^*}
\frac{1}{[z_0^2+\rho_c^2]^3[(x-z_0)^2+\rho_c^2]^3}\ ,
\end{equation}
while the three-body exactly vanishes.

The Borel transform of the correlator for A for arbitrary values
of the parameters $f$ and $t$  has a rather complicated form and
has to be calculated numerically. For the B state it is simple and
the proportional to the two body nucleon one
\begin{equation}
\Pi^B_2(M)_{inst}= -\frac{ 11n_{eff} \langle
\bar{q}q\rangle_M\rho_c^4}{4\cdot1728m_u^*m_s^*}
M^6e^{-z^2/2}\bigg(K_0(z^2/2)+K_1(z^2/2)\bigg), \label{B}
\end{equation}

\begin{figure}[htb]
\begin{minipage}[c]{7cm}
\vspace*{0.5cm}
\psfig{file=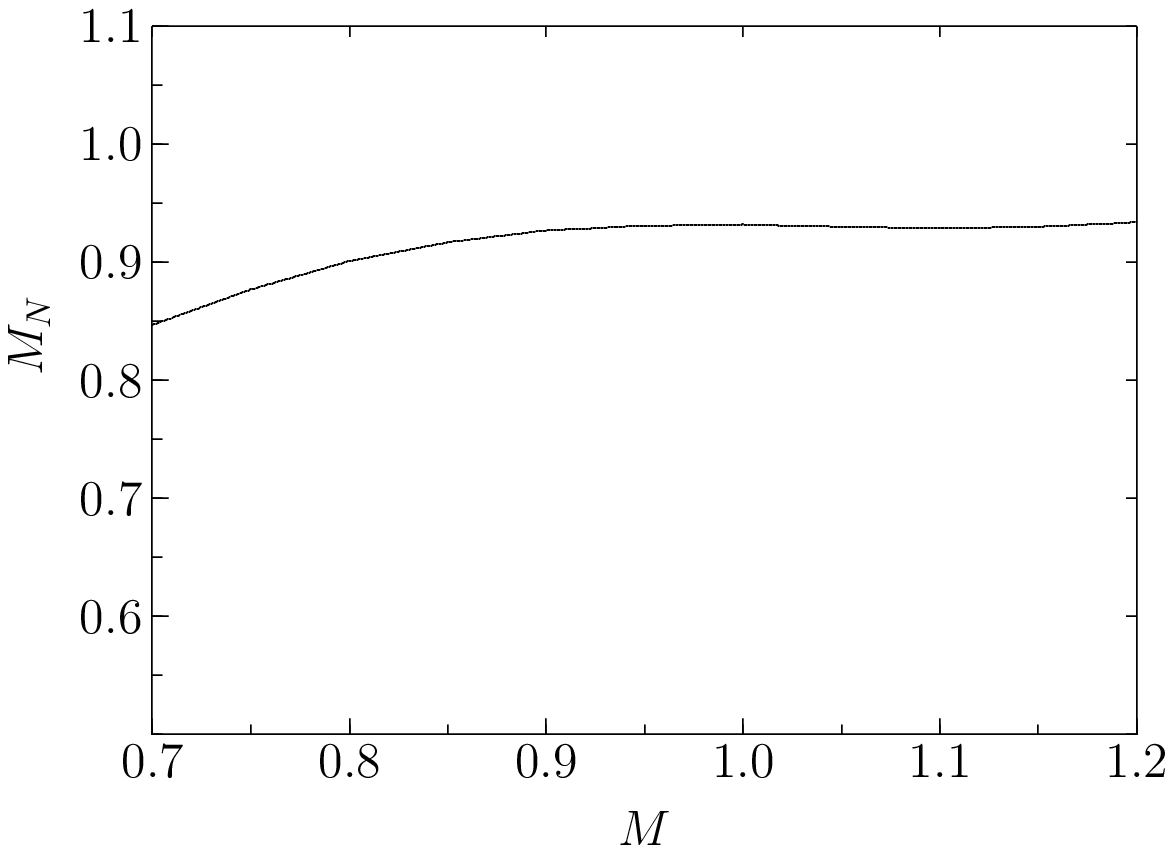,width=7cm,height=6cm}
\caption{\it The  nucleon mass as a function of the Borel
parameter M for the chiral odd SR with $s_0=1.75$ GeV. }
\end{minipage}
\hspace*{0.5cm}
\begin{minipage}[c]{7.cm}
\vspace*{0.5cm}
\psfig{file=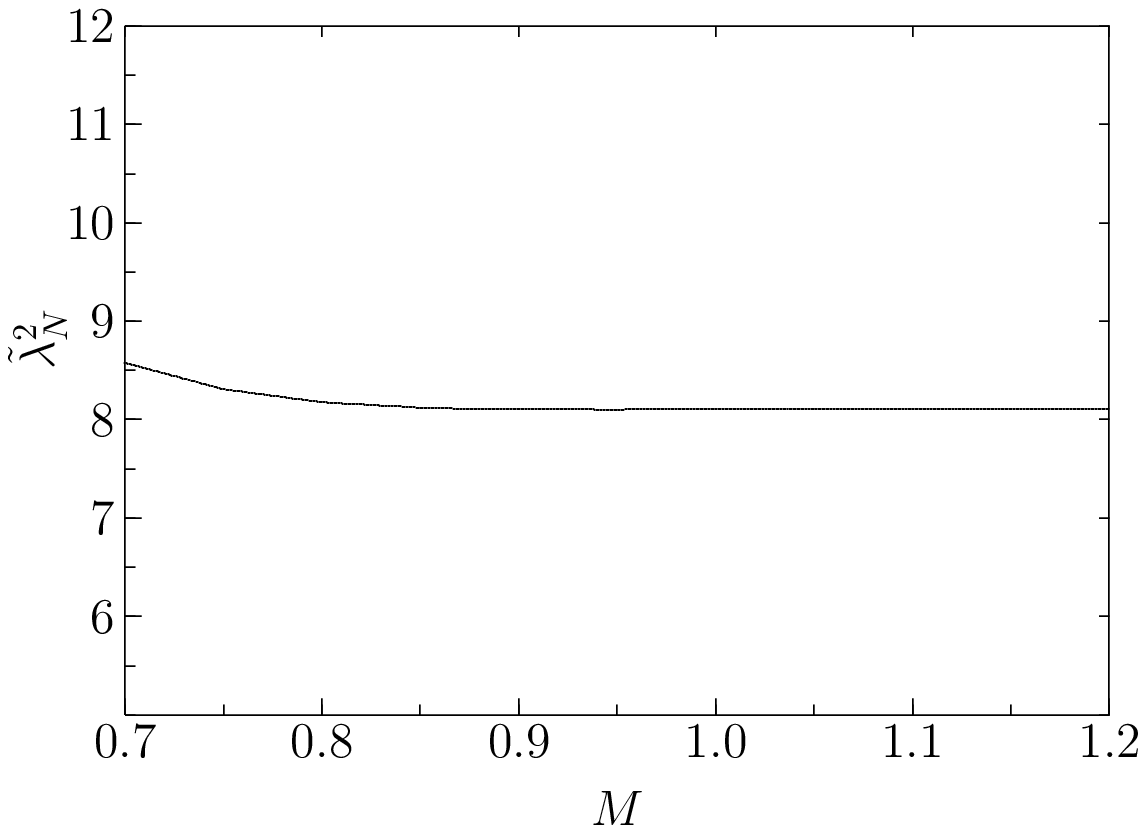,width=7cm,height=6cm}
\caption{\it The  nucleon residue  as a function of the Borel
parameter M for chiral odd SR with $s_0=1.75$ GeV.}
\end{minipage}
\end{figure}

Our estimates of instanton effects make use of the following
relation between the parameters of the Shuryak instanton model
\cite{DEK}
\begin{equation}
\frac{2n_{eff}}{m_q^{*2}}=\frac{3}{2\pi^2\rho_c^2}\ .
\end{equation}
Furthermore, it turns out that, in the model, the size of the
instanton contribution is determined only by value of the average
instanton size in the QCD vacuum $\rho_c$.

Let us discuss first the instanton contribution to the nucleon
case. In the literature there are two slightly different
statements about the effects of instantons on the stability of the
nucleon SR. In ref.\cite{dorkoch} it was shown that the instantons
lead to a significant improvement of the stability of the chiral
odd $\Pi_2$ SR and do not improve the stability of chiral even
$\Pi_1$ SR. However, in ref.\cite{forkel} it was argued that the
instanton contribution (\ref{nucleon1}) also leads to the
appearance of a stability plateau as a function of Borel mass in
the chiral even SR for $f\neq -t$.

Our present exact calculation confirms the results of
ref.~\cite{dorkoch} and gives rise to a stability plateau  for the
chiral odd SR (Figs. 4 and 5). We also obtain the experimental
mass of the nucleon, $M_N=940 MeV$, for a reasonable average size
of the instantons, $\rho_c \approx 1.6$ GeV$^{-1}$. The existence
of a stability plateau in $\Pi_2$ SR is not very sensitive to the
values of the parameters of the nucleon current $f$ and $t$.
However, we did not find such a stability plateau for the chiral
even SR for any choice of the current parameters $f$ and $t$. We
should lastly mention that for the Ioffe current $f=-t$, the
instanton contribution to chiral even SR vanishes explicitly
(recall Eq.({\ref{nucleon1})). Therefore chiral even SR are not
considered in our discussion below.

\begin{figure}[htb]
\begin{minipage}[c]{7cm}
\vspace*{0.5cm}
\psfig{file=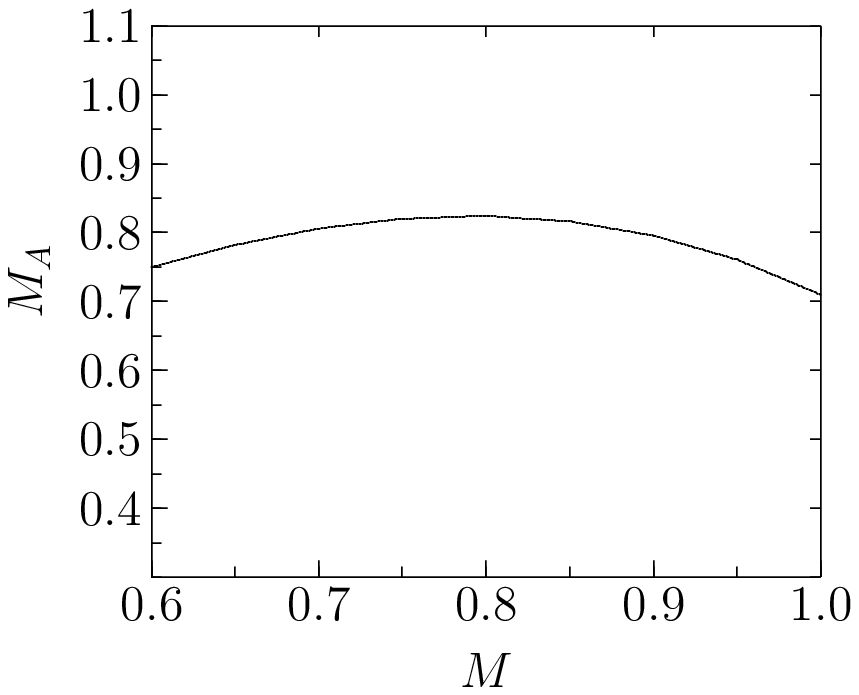,width=7cm,height=6cm}
\caption{\it The A  state mass obtained incorporating the
instanton contributions as a function of the Borel parameter for
chiral odd SR with $s_0=1.8$ GeV. }
\end{minipage}
\hspace*{0.5cm}
\begin{minipage}[c]{7cm}
\vspace*{0.5cm} \psfig{file=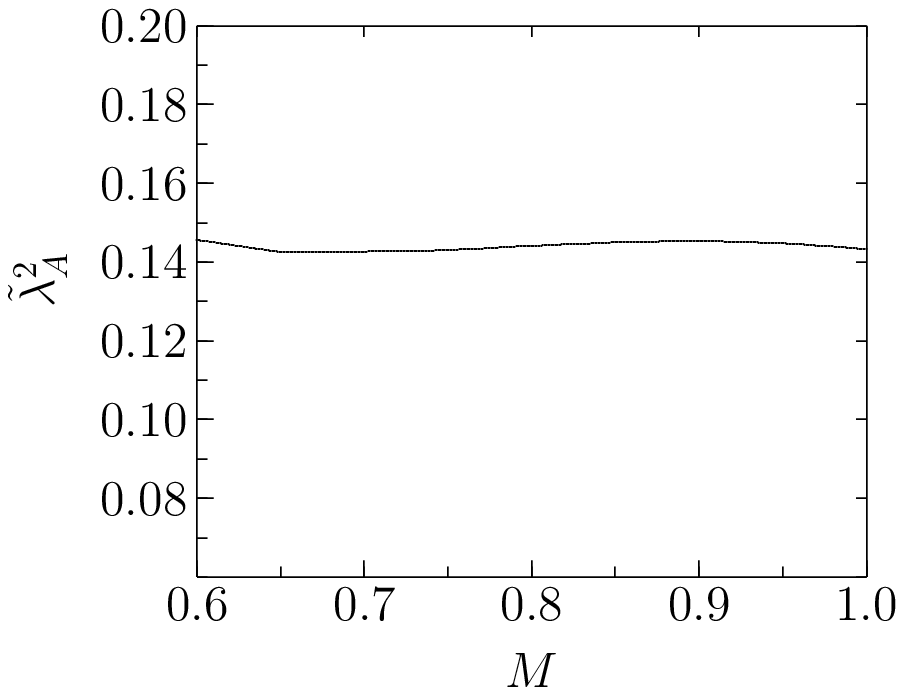,width=7cm,height=6cm}
\caption{\it The A state residue obtained including the instanton
contributions as a function of Borel parameter for chiral odd SR
with $s_0=1.8$ GeV.}
\end{minipage}
\end{figure}

\begin{figure}[htb]
\begin{minipage}[c]{7cm}
\vspace*{0.5cm}
\psfig{file=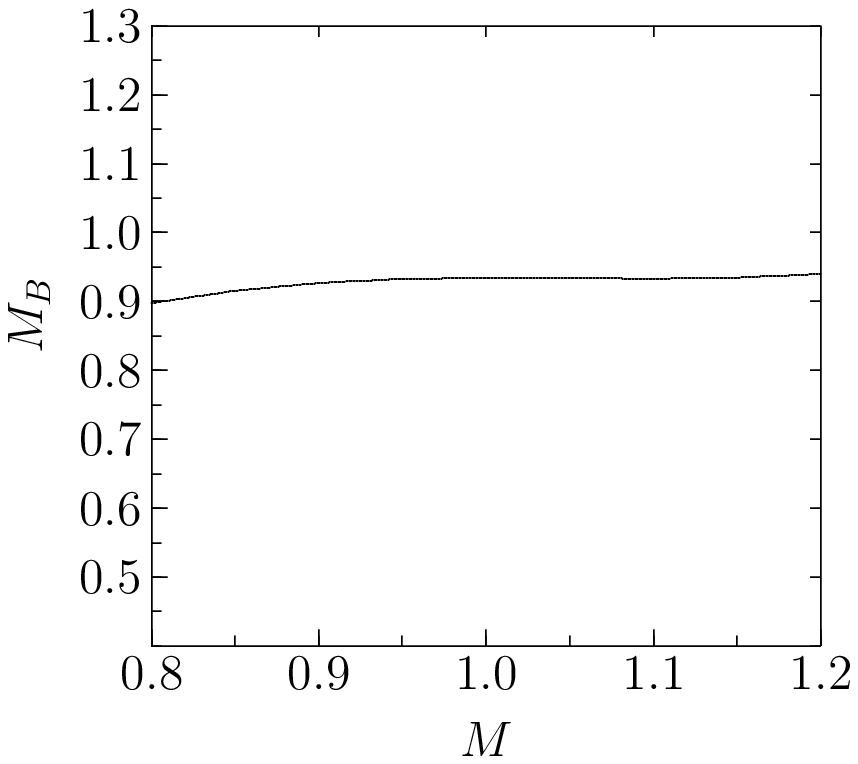,width=7cm,height=6cm}
\caption{\it The B state mass containing the instanton
contributions as a function of Borel parameter M from chiral odd
SR with $s_0=1.72$ GeV. }
\end{minipage}
\hspace*{0.5cm}
\begin{minipage}[c]{7cm}
\vspace*{0.5cm} \psfig{file=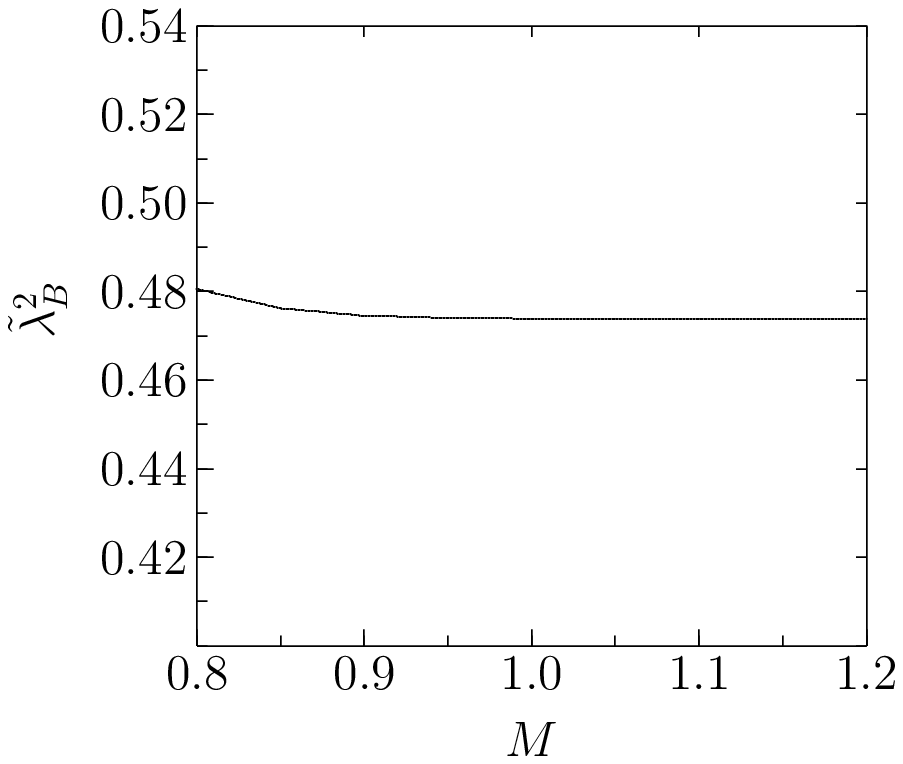,width=7cm,height=6cm}
\caption{\it The B  state residue incorporating the instanton
contributions as a function of Borel parameter M for chiral odd SR
with $s_0=1.72$ GeV.}
\end{minipage}
\end{figure}

The study of the SR provides us with a very light negative parity
A triquark state whose mass is $M_{tri}^A\approx 800$ MeV  for a
mixed A type current with  $f \approx -t$. This state shows a
stability plateau as function of Borel parameter for both mass and
residue (Figs. 6 and 7). $\lambda_A^2$ is positive which insures
that the parity is negative. For the B current we also find  a
negative parity state again with a stability plateau for both mass
and residue (Figs. 8 and 9). The value of its  mass is
$M_{tri}^B\approx 930$ MeV and again $\lambda_B^2$ is positive.

It should be emphasized that without the contribution of the
instantons our analysis of the chiral odd triquark SR would
have shown an absence of stability plateau for both A and B
states. Therefore their mass would have been difficult to
determine.

Our calculation shows that three-body contribution for triquark A
state is very small and vanishes for a Ioffe type triquark current
($f=-t$) as well as for the B current. Therefore, we expect that
three-body instanton induced forces do not play a significant role
in multiquark systems. This conclusion is in agreement with the
result of the calculation of the three-body instanton contribution
to the mass of the $H$-dibaryon within a bag model \cite{DK}.

Recalling the investigations with diquarks
\cite{shuryak3,faccioli} and at the light of our present results
it becomes natural to consider a model for a light pentaquark as
an A--B (mixed) triquark--($ud$) diquark system, with a coupling
between the clusters with non-zero orbital momentum $L=1$
\cite{klv}. In this case centrifugal barrier will suppress quark
rearrangement between the two color clusters. Furthemore, an
additional orbital excitation energy $\approx 400$ MeV (see
\cite{klv}) will bring the total mass of pentaquark to its
observed value. The heavier pentaquark
B--A(mixed)--triquark--($ud$) diquark orthogonal system is
expected to have a mass about $200$ MeV higher. Due to negative
internal parity of the light triquark state the total parity of
pentaquark system in this case  is positive in agreement with the
expectation of the soliton model \cite{chiral}.

\section{Conclusion}

The dynamics of correlated quarks is being appreciated in many
areas of hadron physics. The``wishful" discovery of the $\Theta^+$
and its immediate consequences on the spectrum would allow the
study of multiparticle correlations in QCD in a natural scenario.
In the undesirable circumstances that the pentaquarks, and other
exotics, do not exist the study of quark correlations in other
domains of hadron physics will open up the possibility of further
understanding the dynamics of QCD. The aim of this presentation
has been to single out the importance of the instantons in the
multiparticle dynamics of QCD.

In order to do so we have incorporated in a traditional OPE
calculation of SR the direct instanton effects for triquark
$ud\bar s $ color clusters. We have shown that instantons lead to
a large stability for the correlator of the color triquark current
as a function of the Borel parameter. We observe the formation of
two negative parity $ud \bar s$ states with spin one-half and
isospin zero. These particular triquark states \cite{klv} might be
behind the unusual properties of the observed pentaquark state and
support the Karliner and Lipkin triquark-diquark clusterization
scheme \cite{kl}. We emphasize that all published calculations of
masses of pentaquark within QCD sum rules \cite{ioffe},
\cite{SRpentaquark} should be reanalyzed including the direct
instanton contributions.

We hope that our investigation inspires the study of quark
correlations using lattice theory, a theoretical support to prove
the existence of exotics, and the role played by the instantons in
their dynamics, using the appropriate techniques \cite{Negele}.
Finally, it is clear that the $\Theta^+$ has become now, above
all, an experimental issue which will be solved in the near
future, but our study indicates that correlations are a
consequence of the way we understand QCD dynamics and we hope to
inspire the search for other experimental scenarios where they
might play an important role.

\begin{center}
{\bf Acknowledgments}
\end{center}
We are grateful to A.E.Dorokhov, S.V. Esaibegian and A. Di Giacomo
for very useful discussions. NIK is grateful to the Universities
of Valencia and Pisa  for the warm hospitality. Part of this work
was done under Valencia University - JINR and INFN - JINR
agreements. HJL is grateful to Bogoliubov Laboratory of
Theoretical Physics of JINR for the hospitality during his visit
and to Mannque Rho for constructive comments. Hee-Jung Lee is a
Postdoctoral fellow from SEEU-SB2002-0009. This work was supported
by grants MCyT-FIS2004-05616-C02-01 and GV-GRUPOS03/094~(VV), and
RFBR-03-02-17291, RFBR-04-02-16445 (NIK).

\end{document}